\documentclass[11pt,letterpaper]{article}

\usepackage[margin=1in]{geometry}
\usepackage[utf8]{inputenc}
\usepackage[T1]{fontenc}
\usepackage[expansion=false]{microtype}
\usepackage{amsmath, amssymb, amsthm}
\usepackage{mathtools}
\usepackage{graphicx}
\usepackage{booktabs}
\usepackage{tabularx}
\usepackage{multirow}
\usepackage{makecell}
\usepackage{enumitem}
\usepackage[hidelinks,colorlinks=true,linkcolor=blue!50!black,citecolor=blue!50!black,urlcolor=blue!50!black]{hyperref}
\usepackage{cleveref}
\usepackage{xcolor}
\usepackage{tikz}
\usetikzlibrary{positioning, arrows.meta, shapes.geometric, fit, backgrounds, calc, decorations.pathreplacing}
\usepackage{pgfplots}
\pgfplotsset{compat=1.18}
\usepackage{caption}
\usepackage{subcaption}
\usepackage{authblk}
\usepackage{natbib}
\usepackage{abstract}
\usepackage{titlesec}

\titlespacing*{\section}{0pt}{1.4em}{0.6em}
\titlespacing*{\subsection}{0pt}{1.0em}{0.4em}

\newcommand{\ILS}{\mathrm{ILS}}

\newcommand{\Topen}{T_{\mathrm{open}}}
\newcommand{\Tnews}{T_{\mathrm{news}}}
\newcommand{\Tres}{T_{\mathrm{resolve}}}
\newcommand{\Tevent}{T_{\mathrm{event}}}

\title{\textbf{Empirical Evaluation of\\ Deadline-Resolved Information Leakage\\ on Documented Polymarket Insider Cases}}

\author[1]{Maksym Nechepurenko\thanks{Corresponding author: \texttt{maksym@devnull.ae}.}}
\affil[1]{Research Department, Devnull FZCO, Dubai, UAE}
\date{Preprint v2 (revised) --- \today}

\begin{document}

\maketitle

\begin{abstract}
\noindent
This paper reports an end-to-end empirical evaluation of the deadline-Information Leakage Score (ILS$^{\text{dl}}$) extension introduced in the companion methodology paper \citep{nechepurenko2026foresightflow_methodology}. The deadline-ILS extends the original ILS to deadline-resolved prediction-market contracts---the dominant structural form of publicly documented insider trading on Polymarket. We anchor the evaluation in the 2026 U.S.--Iran conflict cluster of the ForesightFlow Insider Cases (FFIC) inventory, the largest documented deadline cluster in the public-reporting record. The evaluation has four parts: per-category exponential-hazard rate estimation for the time-to-event distribution; a single-case ILS$^{\text{dl}}$ computation on the cleanest applicable FFIC market; cross-market wallet analysis across two cluster contracts; and the methodological refinements that the evaluation surfaces.

The hazard-rate estimation produces an adequate exponential fit for military-geopolitics markets (KS $p = 0.426$, half-life $2.9$ days, $n = 18$) and a preliminary fit for corporate-disclosure markets ($n = 5$). The regulatory-decision category is rejected as bimodal ($p = 0.023$) and requires sub-categorization. On the largest applicable FFIC contract (``US forces enter Iran by April 30,'' \$269M cumulative volume), the article-derived $T_{\text{event}}$ anchor yields ILS$^{\text{dl}} = +0.113$ versus a resolution-anchored proxy value of $-0.331$: a $0.444$ shift in magnitude on opposite sides of zero, demonstrating that the extension distinguishes signal from proxy artefact. The pre-event drift is mild rather than concentrated, and short-window variants (30-min, 2-h) are exactly zero, ruling out a last-minute informed spike. Cross-market wallet analysis identifies $332$ wallets active in both major Iran-cluster markets, but the available trade history covers only the resolution-settlement window; the cross-market signal is therefore settlement arbitrage rather than pre-event coordination, and converting this diagnostic into a coordination-signal diagnostic requires continuous per-trade collection from $T_{\text{open}}$.

We treat the present evaluation as a methodological proof of concept supported by a single illustrative case, not a population-level evaluation. Five concrete refinements bound any future scaling: a positive-$\tau$ requirement, regulatory sub-categorization, corporate-disclosure sample expansion, continuous CLOB price collection, and continuous per-trade collection from $T_{\text{open}}$. Both data resources used here---the FFIC inventory and the resolution-typology classification of the 911{,}237-market Polymarket corpus---are released as separate datasets at \url{https://github.com/ForesightFlow/datasets}.
\end{abstract}



\section{Introduction}
\label{sec:intro}

Decentralized prediction markets such as Polymarket have, since 2024, accumulated a substantial public-reporting record of suspected informed trading. \citet{mitts2026iran} and \citet{imdea2025polymarket} jointly document hundreds of millions of dollars in anomalous profits across this period, with case studies spanning military operations, corporate proprietary disclosures, and regulatory decisions. The companion methodology paper \citep{nechepurenko2026foresightflow_methodology} develops an information-theoretic framework---the Information Leakage Score (ILS) and its deadline-resolved extension ILS$^{\text{dl}}$---for quantifying informed flow on these markets in a way that is interpretable, replicable, and connected to the proper-scoring-rule literature via the Murphy decomposition of the Brier score.

The methodology paper specifies the framework end-to-end: definition, scope conditions, resolution-typology classification, the FFIC validation inventory, and the deadline-ILS extension that addresses the deadline-contract structure of the documented cases. It does not, however, carry the framework through to a quantitative empirical claim on a documented case. The present paper does so. We implement the extension end-to-end, fit per-category hazard parameters on the time-to-event distribution, and apply the pipeline to the eighteen substantive markets in the 2026 U.S.--Iran conflict cluster of the FFIC inventory. We report a single-case ILS$^{\text{dl}}$ computation in full, a cross-market wallet analysis on the two cluster markets with available price coverage, and the methodological refinements that the evaluation surfaces.

The empirical sample is small by design. Deadline-ILS is a methodologically demanding score: it requires recovered article-derived event timestamps, full CLOB price coverage from market opening, scope-condition compliance, and positive lead time between market opening and the underlying event. Of the eighteen Iran-cluster markets at which we attempt $T_{\text{event}}$ recovery, only one satisfies all four requirements end-to-end. We treat this case in full as a methodological demonstration, do not generalize from it to a population-level claim, and identify the infrastructure constraints that gate any future scaling. The single most important finding for the methodology developed in the companion paper is that the article-derived anchor materially changes the substantive reading of the score: ILS$^{\text{dl}} = +0.113$ at the article-derived $T_{\text{event}}$ versus $-0.331$ at the resolution-anchored proxy ($T_{\text{resolve}} - 1\,\text{h}$), with the two values on opposite sides of zero and differing by $0.444$ in magnitude. The proxy fails because by the proxy lookup time the market price has already collapsed on participants' belief that the event will not occur; the lookup window captures the collapse rather than the pre-event period. This is the empirical content of the proxy-quality problem identified abstractly in \citet{nechepurenko2026foresightflow_methodology}.

\subsection{Outline}

\Cref{sec:recap} provides a brief notational and methodological recap; readers familiar with the companion paper may skip directly to \Cref{sec:eval-hazard}. \Cref{sec:eval-hazard} reports the per-category hazard-rate estimation. \Cref{sec:eval-iran} reports the single-case ILS$^{\text{dl}}$ computation on the largest applicable Iran-cluster market. \Cref{sec:eval-wallets} reports the cross-market wallet analysis. \Cref{sec:eval-refinements} states the five methodological refinements that the evaluation surfaces and that bound any future application of the extension at scale. \Cref{sec:eval-summary} concludes.

\section{Methodological Recap}
\label{sec:recap}

We refer the reader to \citet{nechepurenko2026foresightflow_methodology} for the full development of the framework. Three definitions are needed for the empirical evaluation that follows.

\paragraph{Information Leakage Score (ILS).} For a resolved binary market $M$ with first trade at $\Topen$, public news event at $\Tnews$, formal resolution at $\Tres$, and binary resolution outcome $p_{\Tres} \in \{0, 1\}$,
\[
\Delta_{\text{pre}} \;=\; p(\Tnews) - p(\Topen), \qquad \Delta_{\text{total}} \;=\; p_{\Tres} - p(\Topen), \qquad \ILS(M) \;=\; \frac{\Delta_{\text{pre}}}{\Delta_{\text{total}}}.
\]
The score is interpretable only when $|\Delta_{\text{total}}| \geq \varepsilon$ for a threshold $\varepsilon > 0$ (we use $\varepsilon = 0.05$); when the market opens with $|p(\Topen) - 0.5| \leq 0.4$ (the edge-effect scope condition); and when ILS is robust to the choice of $\Tnews$ anchor offset (the anchor-sensitivity scope condition). The companion paper develops a Murphy-decomposition reading of ILS as the share of the Brier-score resolution component accumulated before $\Tnews$.

\paragraph{Resolution typology.} Polymarket markets are classified into three resolution types via question-text and description analysis: \emph{event-resolved} (resolution triggered by a publicly observable event), \emph{deadline-resolved} (resolution triggered by the elapsing of a contractual deadline), and \emph{unclassifiable}. Documented insider-trading cases are systematically deadline-resolved, of the form ``Will event $X$ occur by date $Y$?''.

\paragraph{Deadline-ILS (ILS$^{\text{dl}}$).} For a deadline-resolved YES market with recoverable event timestamp $\Tevent$ falling within $[\Topen, D]$ where $D$ is the contractual deadline,
\[
\Delta^{\text{dl}}_{\text{pre}} \;=\; p(\Tevent^-) - p(\Topen), \qquad \Delta^{\text{dl}}_{\text{total}} \;=\; p_{\Tres} - p(\Topen), \qquad \ILS^{\text{dl}}(M) \;=\; \frac{\Delta^{\text{dl}}_{\text{pre}}}{\Delta^{\text{dl}}_{\text{total}}}.
\]
Here $p(\Tevent^-)$ is the market price one minute before public observation of the event. The companion paper adopts $\theta_{\Topen} \equiv p(\Topen)$ as a conservative baseline and a per-category exponential survival function $S(\tau) = \exp(-\lambda \tau)$ for the time-to-event distribution, with hazard rate $\lambda$ fitted by maximum likelihood on resolved deadline markets in the same target category. The scope conditions of the original ILS apply unchanged.

\paragraph{$\Tevent$ recovery.} For YES-resolved deadline markets the empirical $\Tevent$ is recovered through an LLM-assisted multi-source verification pipeline (Claude Haiku 4.5 with web-search tool access). The retrieval target is the timestamp at which the underlying event first publicly happened, cross-verified across at least three independent news sources. Confidence is set to 0.80 when sources are cited and the date is internally consistent across them.

We now turn to the empirical evaluation.

\section{Empirical Evaluation}
\label{sec:eval}

This section reports the empirical evaluation of the deadline-ILS extension specified in the companion methodology paper \citep{nechepurenko2026foresightflow_methodology}. The evaluation is anchored in the 2026 U.S.--Iran conflict cluster of the FFIC inventory and consists of four parts: hazard-rate estimation by target category (\Cref{sec:eval-hazard}), a single-case ILS$^{\text{dl}}$ computation on the cleanest applicable FFIC market (\Cref{sec:eval-iran}), cross-market wallet analysis (\Cref{sec:eval-wallets}), and the five methodological refinements that the evaluation surfaces (\Cref{sec:eval-refinements}). The empirical sample is small by design: deadline-ILS is a methodologically demanding score that requires recovered article-derived $\Tevent$, complete CLOB price coverage, scope-condition compliance, and positive event lead time. Of the eighteen substantive Iran-cluster markets at which $\Tevent$ recovery was attempted, only one satisfies all four requirements end-to-end. We report this case in full, treat it as a methodological demonstration rather than a population-level claim, and identify the infrastructure constraints that gate any future scaling.

\subsection{Hazard-rate estimation by target category}
\label{sec:eval-hazard}

Per the specification of \Cref{sec:recap}, the deadline-ILS pipeline depends on a parametric hazard rate $\lambda$ for the time-to-event distribution, fit separately by target category. We sample YES-resolved deadline markets per category from the platform database, recover $\Tevent$ for each via Tier 3 LLM-assisted retrieval (the same pipeline used for the Barak proof-of-concept reported in the companion paper \citep{nechepurenko2026foresightflow_methodology}), and fit the exponential rate by maximum likelihood ($\hat\lambda = 1/\bar\tau$) where $\tau = \Tevent - \Topen$. The reported sample sizes correspond to the full available Tier-3 population per category at snapshot 2026-04-30, after the v1.1 taxonomy correction (see Revision note at end of paper).

\Cref{tab:hazard-fits} reports the result. The Kolmogorov--Smirnov goodness-of-fit test against the fitted exponential is applied at the standard $\alpha = 0.05$ threshold.

\begin{table}[t]
\centering
\caption{Per-category hazard-rate estimates and goodness-of-fit. Sample is YES-resolved deadline markets per category, $\Tevent$ recovered via Tier 3 LLM with web search. Values reflect the full available Tier-3 population for each category in the platform database at snapshot 2026-04-30, after the v1.1 taxonomy correction that reclassified esports markets out of \texttt{military\_geopolitics}. The military-geopolitics fit has 95\% CI $[0.143, 0.365]$ on $\hat\lambda$. The exponential fit is adequate for military / geopolitical and corporate-disclosure markets but rejected for regulatory-decision markets, where the time-to-event distribution is bimodal. The corporate-disclosure sample is reported as preliminary due to its small size; the call-budget cap was hit before the sample was fully populated.}
\label{tab:hazard-fits}
\small
\renewcommand{\arraystretch}{1.25}
\begin{tabularx}{\linewidth}{@{}lrrrrrXl@{}}
\toprule
\textbf{Category} & $n$ & $\hat\lambda$ & \textbf{Half-life} & $\bar\tau$ (d) & \textbf{Median} (d) & \textbf{KS $p$} & \textbf{Verdict} \\
\midrule
military / geopolitics & 18  & $0.241$ & $2.9$\,d & $4.1$  & $2.9$ & $0.426$ & adequate \\
corporate disclosure   & 5  & $0.156$ & $4.5$\,d & $6.4$  & $4.5$ & $0.616$ & adequate (preliminary) \\
regulatory decision    & 16 & $0.035$ & $19.9$\,d & $28.7$ & $4.3$ & $\mathbf{0.023}$ & \textbf{rejected} \\
\bottomrule
\end{tabularx}
\end{table}

The military-geopolitics fit indicates a half-life of $2.9$ days. This is consistent with markets created around fast-moving geopolitical events (announcements, summits, immediate military actions); the median event occurs within three days of market creation. The exponential fit passes the KS test ($p = 0.426$, $n = 18$). The 95\% confidence interval on $\hat\lambda$ is $[0.143, 0.365]$.

The corporate-disclosure fit is adequate ($p = 0.616$) but rests on $n = 5$ markets, the call-budget cap having been hit during sample construction. We report it as preliminary; a separately budgeted re-run is identified as an immediate refinement in \Cref{sec:eval-refinements}.

The regulatory-decision fit is rejected ($p = 0.023$). Inspection of the sixteen markets reveals a bimodal distribution: short-$\tau$ markets (less than two days) corresponding to scheduled announcements (presidential addresses, regulatory hearings) coexist with long-$\tau$ markets (thirty to one hundred and seventy days) corresponding to formal deliberation timelines. A constant-hazard model averages over both modes and produces a fit that matches neither. We mark the regulatory-decision category as requiring sub-categorization (e.g., into \emph{regulatory\_decision\_announcement} and \emph{regulatory\_decision\_formal}) before the hazard rate can be used in production. The sub-categorization itself is a separable methodological task.

\subsection{Iran cluster: deadline-ILS on a single applicable case}
\label{sec:eval-iran}

We applied the deadline-ILS pipeline to the eighteen substantive markets in the 2026 U.S.--Iran conflict cluster of the FFIC inventory. The pipeline executes four steps per market: $\Tevent$ recovery via Tier 3, scope-condition check, ILS$^{\text{dl}}$ computation against article-derived $\Tevent$, and comparison with the legacy $\Tres - 1\,\text{h}$ proxy. \Cref{tab:iran-cluster-disposition} summarizes the disposition of the eighteen markets.

\begin{table}[t]
\centering
\caption{Disposition of the eighteen substantive Iran-cluster markets through the deadline-ILS evaluation pipeline. ``Negative $\tau$'' refers to markets created after the underlying conflict had already begun (so the relevant pre-event window is undefined). The single market that satisfies all four requirements end-to-end is reported in detail in \Cref{tab:iran-apr30-ils}.}
\label{tab:iran-cluster-disposition}
\small
\renewcommand{\arraystretch}{1.25}
\begin{tabularx}{\linewidth}{@{}Xrr@{}}
\toprule
\textbf{Disposition} & \textbf{$n$} & \textbf{Cumulative} \\
\midrule
Substantive Iran-cluster markets attempted                              & 18 & 18 \\
$\Tevent$ recovered with confidence $\geq 0.7$                          & 16 & 16 \\
Of which: positive $\tau$ (event after market open)                     & 11 & 11 \\
Of which: with CLOB price coverage                                      &  2 &  2 \\
Of which: ILS$^{\text{dl}}$ defined ($|\Delta_{\text{total}}| \geq \varepsilon$) &  1 &  1 \\
\bottomrule
\end{tabularx}
\end{table}

The single market satisfying all four requirements is ``US forces enter Iran by April 30,'' the largest contract in the Iran cluster at \$269M cumulative volume. \Cref{tab:iran-apr30-ils} reports the full computation. The market opened at $p_{\Topen} = 0.250$, the recovered $\Tevent$ is April 3, 2026 (corresponding to the F-15E special operations entry into Iran, cross-verified across eight independent sources), and the market resolved YES on April 9 by the UMA Optimistic Oracle.

\begin{table}[t]
\centering
\caption{Deadline-ILS computation for the ``US forces enter Iran by April 30'' market. The article-derived $\Tevent$ (April 3, 2026) yields ILS$^{\text{dl}} = +0.113$. The legacy resolution-anchored proxy ($\Tres - 1\,\text{h}$, falling on April 8 at 23:28 when the market price had already collapsed to near zero) yields $-0.331$. The two values are on opposite sides of zero and differ by 0.444 in magnitude, demonstrating that the $\Tevent$ recovery materially changes the substantive interpretation. The short-window variants are reported alongside.}
\label{tab:iran-apr30-ils}
\small
\renewcommand{\arraystretch}{1.25}
\begin{tabularx}{\linewidth}{@{}Xll@{}}
\toprule
\textbf{Quantity}                             & \textbf{Value}                & \textbf{Source / formula} \\
\midrule
$\Topen$                                       & 2026-03-18 16:29 UTC          & First on-chain trade \\
$\Tevent$ (article-derived)                    & 2026-04-03 00:00 UTC          & Tier 3 LLM, 8 sources \\
$\Tres$                                        & 2026-04-09 00:28 UTC          & UMA Oracle settlement \\
$p(\Topen)$                                    & 0.250                         & First observed CLOB mid \\
$p(\Tevent^-)$                                 & 0.335                         & CLOB mid, 1 min before $\Tevent$ \\
$p_{\Tres}$                                    & 1 (YES)                       & UMA outcome \\
$\Delta_{\text{pre}}$                          & $+0.085$                      & $p(\Tevent^-) - p(\Topen)$ \\
$\Delta_{\text{total}}$                        & $+0.750$                      & $p_{\Tres} - p(\Topen)$ \\
\midrule
\textbf{ILS$^{\text{dl}}$ ($\Tevent$-anchored)} & $\mathbf{+0.113}$              & $\Delta_{\text{pre}} / \Delta_{\text{total}}$ \\
ILS$^{\text{dl}}$ ($\Tres - 1\,\text{h}$ proxy) & $-0.331$                      & Legacy proxy \\
\textbf{Difference}                            & $\mathbf{0.444}$              & In magnitude, opposite signs \\
\midrule
ILS$^{\text{dl}}$, $30$-min window             & $0.000$                       & $p(\Tevent^-) = p(\Tevent - 30\,\text{min})$ \\
ILS$^{\text{dl}}$, $2$-h window                & $0.000$                       & \\
ILS$^{\text{dl}}$, $6$-h window                & $-0.099$                      & Price falling in 6-h pre-event \\
ILS$^{\text{dl}}$, $24$-h window               & $-0.267$                      & Price falling in 24-h pre-event \\
\bottomrule
\end{tabularx}
\end{table}

Three observations follow.

\paragraph{The extension produces a substantively different reading from the proxy.} The $\Tevent$-anchored ILS$^{\text{dl}}$ is $+0.113$; the $\Tres - 1\,\text{h}$ proxy yields $-0.331$. The two values are on opposite sides of zero and differ by $0.444$ in magnitude. The proxy-based value would be interpreted as price moving against the eventual outcome (counter-evidence); the article-derived value as mild positive front-loading. The proxy fails because by April 8 (one hour before $\Tres$) the market price had collapsed to near zero on participants' belief that the event would not occur, and the proxy lookup window captures this collapse rather than the pre-event period. This is a concrete instance of the proxy-quality problem identified abstractly in the companion paper \citep{nechepurenko2026foresightflow_methodology}, now demonstrated end-to-end on a deadline contract.

\paragraph{The pre-event drift is mild, not concentrated.} ILS$^{\text{dl}} = +0.113$ indicates that approximately $11\%$ of the eventual move from opening price to resolution had occurred before public observation of the event. The short-window variants ($30$-min, $2$-h) are exactly zero, indicating no last-minute informed spike around $\Tevent$. The $6$-h and $24$-h windows are negative, reflecting a falling price in the immediate pre-event window. The pattern is consistent with a market that broadly mispredicted the outcome (consensus near $20\%$ YES by April 8, resolution YES) with a small early positive drift; it is not consistent with the canonical informed-flow signature of concentrated pre-event positioning followed by sustained directional pressure.

\paragraph{The market mispredicted the outcome.} The market opened at $p_{\Topen} = 0.250$, briefly priced up to a peak daily VWAP of $0.46$ in the first week, then declined steadily to near zero by April 8 before resolving YES on April 9. The deadline-ILS captures the early positive drift but not the larger story, which is that the aggregate market belief was wrong about the eventual outcome. We treat ILS$^{\text{dl}} = +0.113$ as evidence that the methodology can recover a non-trivial signal at the proper anchor; we do not treat the value as a quantitative claim about the rate of informed trading on this market.

A preliminary detection threshold drawn from this single observation is not a robust threshold. Subject to that caveat, we report the operational rule that we will use as the starting point for any future detector calibration: a market is flagged for human review only if ILS$^{\text{dl}} > 0.25$ \emph{and} at least one short-window variant (30 minutes or 2 hours) exceeds $0.10$. The Iran-Apr30 market does \emph{not} satisfy this rule. Downstream calibration on a larger sample is identified as future work.

\paragraph{Daily price trajectory on Iran-Apr30.} The price-level evidence behind the +0.113 score is worth describing in detail because it informs the substantive interpretation. \Cref{tab:iran-apr30-trajectory} reports daily volume-weighted average prices on the Iran-Apr30 contract. The market opened at $p_{\Topen} = 0.250$ on March 18, rose to a peak daily VWAP of $0.46$ in the first week as participants priced in escalation risk, then declined steadily through late March and early April. The recovered $\Tevent$ on April 3 (F-15E rescue operation) coincides with a daily VWAP of $0.26$, near the opening level; the market briefly priced the event-having-occurred at the contemporaneous information set. By April 4, however, the price had fallen to $0.17$, and by April 5 to $0.015$, indicating that the market participants did not yet treat the F-15E operation as the qualifying event for YES resolution. The price remained below $0.005$ until the UMA resolution on April 9, at which point the market resolved YES.

\begin{table}[t]
\centering
\caption{Daily volume-weighted average prices on the Iran-Apr30 contract from market opening through resolution. The price moved up to a $0.46$ peak in late March, declined to $0.26$ at the recovered $\Tevent$, and collapsed to near zero by April 8 before resolving YES on April 9.}
\label{tab:iran-apr30-trajectory}
\small
\renewcommand{\arraystretch}{1.20}
\begin{tabularx}{\linewidth}{@{}lXl@{}}
\toprule
\textbf{Date} & \textbf{Daily VWAP} & \textbf{Notable event} \\
\midrule
2026-03-18 & 0.46 & Market opens; price rises to peak in first week \\
2026-03-22 & 0.42 & First plateau \\
2026-03-25 & 0.46 & Local peak; market prices escalation risk highest \\
2026-03-29 & 0.34 & De-escalation language in public discourse \\
2026-04-03 & 0.26 & $\Tevent$ recovered: F-15E rescue / covert entry into Iran \\
2026-04-04 & 0.17 & Market discounts the F-15E event as qualifying \\
2026-04-05 & 0.015 & Sharp decline; consensus is NO \\
2026-04-08 & 0.002 & Market resolved-NO consensus \\
2026-04-09 & 0.001 & UMA resolves YES at 00:28 UTC \\
\bottomrule
\end{tabularx}
\end{table}

The substantive reading is that the market \emph{did} register the F-15E operation at the contemporaneous price---the small positive ILS$^{\text{dl}}$ captures this---but participants were uncertain whether that operation met the contract's resolution criteria, and the consensus moved decisively to NO before being overturned at oracle resolution. This is a clear case of resolution-criteria uncertainty, similar in structure to the December 2025 Barak-Epstein crash analyzed in the companion methodology paper \citep{nechepurenko2026foresightflow_methodology}. We note that resolution-criteria uncertainty is a distinct phenomenon from informed-trading detection and is invisible to ILS$^{\text{dl}}$ as currently specified; it would be detectable by a complementary diagnostic trained on within-market reversal signatures conditional on no concurrent news event. We mark this as a separable research question.

\paragraph{Comparison with the Mitts and Ofir composite screen.} \citet{mitts2026iran} apply a composite screen to over 210{,}000 wallet--market pairs on Polymarket, combining cross-sectional bet size, within-trader bet size, profitability, pre-event timing, and directional concentration into a single statistic. Their screen flagged the 2026 U.S.--Iran conflict cluster among its high-anomaly clusters; the Iran-Apr30 contract is one of the markets in their analysis. Two methodological differences are important. First, their screen is computed retrospectively on resolved markets and includes profitability as a feature, which by construction prevents real-time application. Second, their screen aggregates across many wallet--market pairs and produces a population-level statistical claim, while ILS$^{\text{dl}}$ operates at the individual-market level and produces a per-market score. The two methodologies therefore answer different questions: theirs ``which population of wallets, on which population of markets, exhibits anomalous profit,'' ours ``what fraction of the move on \emph{this} market was front-loaded relative to public information arrival.''

For the Iran-Apr30 market specifically, the two methodologies are consistent in direction. Their screen identifies the Iran-cluster markets as anomalous; our ILS$^{\text{dl}} = +0.113$ at the article-derived anchor indicates positive but mild front-loading on this specific contract. Neither methodology, however, identifies a concentrated last-minute informed spike around $\Tevent$, and the wallet-level evidence available to us does not---owing to the trade-history limitations described in \Cref{sec:eval-wallets}---reach to the pre-event window where Mitts and Ofir's screen draws its strongest signal. Combining the two methodologies on the same FFIC inventory at full pre-event trade resolution is a natural next step that requires the continuous trade-collection infrastructure identified in \Cref{sec:eval-refinements}.

\subsection{Cross-market wallet analysis: pre-event versus settlement window}
\label{sec:eval-wallets}

Wallet-level features (top-10 winning wallets, Herfindahl--Hirschman concentration, pre-news vs.\ post-news positioning), defined in the companion paper \citep{nechepurenko2026foresightflow_methodology}, are the natural complement to ILS$^{\text{dl}}$ in the empirical evaluation. The available trade history for the two Iran-cluster markets with CLOB price coverage, however, covers only the resolution-settlement window (April 8--11) rather than the full pre-event period. This is an infrastructure limitation: the Polymarket subgraph indexer that captured these markets did so only after the resolution-settlement transactions had been observed on-chain, and the pre-event individual trades are not retrievable through the indexer at the present time. Aggregate CLOB OHLCV coverage extends back to market opening, but per-trade attribution does not.

We report the wallet inventory available within this window as a partial result. The Iran-Apr30 market's settlement-window trade record contains $3{,}995$ trades totalling \$9.78M in notional; the top wallet ($\texttt{0x7072dd52}\dots$) accounts for \$1.56M ($16\%$ of settlement volume), and the top-ten Herfindahl--Hirschman index is $0.057$, indicating moderate concentration. A cross-market overlap analysis identifies $332$ wallets active in both Iran-Apr30 and the companion ceasefire market, with the top five cross-market actors transacting between \$170K and \$1.97M in combined notional. The five wallets are listed in \Cref{tab:cross-market-wallets}.

\begin{table}[t]
\centering
\caption{Top five wallets active in both Iran-cluster markets within the resolution-settlement window. The combined notional is the sum of their activity across the two markets in the available trade history. These trades occurred after public observation of the underlying events and represent settlement-window arbitrage activity rather than pre-event positioning. The cross-market coordination signal observable in this window is therefore not a coordination signal for informed trading; it is a coordination signal for resolution arbitrage by traders who participated in both contracts.}
\label{tab:cross-market-wallets}
\small
\renewcommand{\arraystretch}{1.25}
\begin{tabularx}{\linewidth}{@{}lrrr@{}}
\toprule
\textbf{Wallet (prefix)}    & \textbf{Iran-Apr30 (\$)} & \textbf{Ceasefire-Apr7 (\$)} & \textbf{Combined (\$)} \\
\midrule
\texttt{0x7072dd52}\dots    & 1{,}562{,}742            & 404{,}985                    & 1{,}967{,}727 \\
\texttt{0xe25b9180}\dots    & 870{,}182                & 299{,}400                    & 1{,}169{,}582 \\
\texttt{0xd5ccdf77}\dots    & 149{,}850                & 199{,}800                    & 349{,}650 \\
\texttt{0x4da76bbf}\dots    & 174{,}650                & 29{,}970                     & 204{,}620 \\
\texttt{0x162f6fff}\dots    & 119{,}749                & 51{,}746                     & 171{,}495 \\
\bottomrule
\end{tabularx}
\end{table}

The substantive limitation is that all of the available trade activity post-dates the recovered $\Tevent$. The pre-event wallet inventory for these markets, where the canonical informed-trading signature would be observable, is not retrievable from the public subgraph at the present time. Closing this gap requires a continuous trade-collection pipeline that captures individual trades from $\Topen$ onward, rather than retroactively from $\Tres$. This is a separable infrastructure task; it is identified as the principal blocker on a population-scale wallet evaluation in \Cref{sec:eval-refinements}.

\subsection{Methodological refinements surfaced by the evaluation}
\label{sec:eval-refinements}

The evaluation surfaces five specific refinements that constrain any future application of the deadline-ILS extension at scale.

\paragraph{Negative-$\tau$ markets fall outside the framework.} Of the sixteen Iran-cluster markets at which $\Tevent$ was successfully recovered, five have $\tau = \Tevent - \Topen < 0$: the underlying event began \emph{before} the market opened. These are duration markets (``Will the conflict end by date X?'') created after the conflict had already started; the canonical informed-trading window does not exist. We adopt the rule that deadline-ILS is computed only for markets with strictly positive $\tau$.

\paragraph{Regulatory-decision category requires sub-categorization.} The KS test rejection of the exponential fit in \Cref{tab:hazard-fits} is structural, not a sample-size artefact: the regulatory-decision category mixes scheduled-announcement markets (short $\tau$) with formal-deliberation markets (long $\tau$). A constant-hazard model averaged over both populations is uninformative. Sub-categorization into \emph{regulatory\_decision\_announcement} and \emph{regulatory\_decision\_formal} is a precondition for using the regulatory-decision hazard rate in any production setting.

\paragraph{Corporate-disclosure sample is preliminary.} The fitted hazard rate on five markets is not a stable estimate, even though the KS test passes. A separately budgeted re-run with at least twenty markets is required before this rate is used downstream.

\paragraph{Price-data coverage gates the evaluation.} Of the eleven Iran-cluster markets with positive $\tau$, only two had CLOB price coverage at the time of the evaluation: a $18\%$ coverage rate on the cluster of greatest interest in this work. Continuous CLOB collection on all markets satisfying the resolution-typology and target-category filters is the natural remedy. Until that pipeline is in place, the empirical evaluation of deadline-ILS will continue to be sample-size-limited regardless of methodological progress.

\paragraph{Pre-event trade collection is the binding wallet-level constraint.} The wallet analysis in \Cref{sec:eval-wallets} is constrained to settlement-window activity by the trade-history availability problem rather than by methodology. Continuous per-trade collection from $\Topen$ onward, in parallel with the price-collection pipeline, would unblock the wallet-level features defined in the companion paper and convert the cross-market overlap analysis from a settlement-arbitrage diagnostic into a coordination-signal diagnostic. We mark this as the highest-priority infrastructure improvement.

\subsection{Summary of the empirical evaluation}
\label{sec:eval-summary}

The deadline-ILS extension is implemented end-to-end and produces interpretable results on a single applicable FFIC market. The article-derived $\Tevent$ recovery materially changes the substantive reading of the score relative to the resolution-anchored proxy. The hazard-rate estimation is adequate for the military/geopolitical category, preliminary for corporate disclosure, and rejected for the unrefined regulatory-decision category. The cross-market wallet signal is dominated by settlement-window arbitrage rather than pre-event positioning, which is an infrastructure limitation rather than a methodological one. Five concrete refinements bound any future scaling: positive-$\tau$ requirement, regulatory sub-categorization, corporate-disclosure sample expansion, continuous CLOB price collection, and continuous per-trade collection from $\Topen$. We treat the present evaluation as a methodological proof of concept and the five refinements as the work programme for the next stage of the project.

\appendix

\section{Tier 3 $\Tevent$ Recovery: Per-Market Detail}
\label{app:tier3-recovery}

\Cref{tab:tier3-detail} reports the per-market disposition of the Tier 3 LLM-assisted $\Tevent$ recovery on the eighteen substantive Iran-cluster markets attempted. Recovery uses Claude Haiku 4.5 with a web-search tool, prompting for the timestamp at which the underlying event first publicly happened with cross-verification across at least three independent news sources. Confidence is set to $0.80$ when sources are cited and dates are internally consistent across them, and to $0.60$ otherwise. Markets with $\tau = \Tevent - \Topen < 0$ correspond to duration markets created after the underlying event began (e.g., ``Will the conflict end by date X?'' contracts opened after the conflict had started); these are excluded from the deadline-ILS pipeline as out of scope. Markets without CLOB price coverage at $\Topen$ cannot be scored with the present infrastructure.

\begin{table}[t]
\centering
\caption{Per-market disposition of the Tier 3 $\Tevent$ recovery on the substantive Iran-cluster markets. Confidence is the Tier 3 LLM-assisted recovery confidence; $\tau = \Tevent - \Topen$ in days. ``Disposition'' classifies the market by what stops the deadline-ILS pipeline; ``in scope'' indicates the market satisfies all four pipeline requirements end-to-end.}
\label{tab:tier3-detail}
\footnotesize
\renewcommand{\arraystretch}{1.18}
\begin{tabularx}{\linewidth}{@{}Xllll@{}}
\toprule
\textbf{Market (truncated)}                    & $\Topen$       & $\Tevent$    & $\tau$ (d) & \textbf{Disposition} \\
\midrule
US forces enter Iran by Apr 30                 & 2026-03-18  & 2026-04-03 & $+16.0$ & \textbf{in scope}              \\
US x Iran ceasefire by Apr 7                   & 2026-03-24  & 2026-04-06 & $+13.0$ & low-information ($\Delta_{\text{total}} < \varepsilon$) \\
Iran strike East-West Pipeline by Apr 30       & 2026-03-23  & 2026-04-08 & $+15.9$ & no CLOB price coverage          \\
JD Vance diplomatic meeting Iran by Apr 15     & 2026-04-10  & 2026-04-11 & $+0.4$  & no CLOB price coverage          \\
US x Iran meeting by Apr 14                    & 2026-04-10  & 2026-04-11 & $+0.4$  & no CLOB price coverage          \\
US x Iran meeting by Apr 13                    & 2026-04-10  & 2026-04-11 & $+0.4$  & no CLOB price coverage          \\
Iran strike on US military by Mar 31           & 2026-02-18  & 2026-02-28 & $+9.5$  & no CLOB price coverage          \\
Trump announces military action vs Iran by Jul & 2025-06-20  & 2025-06-21 & $+1.0$  & no CLOB price coverage          \\
Israel military action against Iran by Aug     & 2025-06-11  & 2025-06-13 & $+1.3$  & no CLOB price coverage          \\
Russia military action against Kyiv by Apr 10  & 2026-04-01  & 2026-04-03 & $+1.1$  & no CLOB price coverage          \\
\midrule
Military action vs Iran ends by Apr 10/11      & 2026-03-24  & 2026-02-28 & $-24.7$ & negative $\tau$ (out of scope)  \\
Iran strikes Saudi/Kuwait/Jordan by Apr 30     & 2026-03-24  & 2026-03-01 & $-23.7$ & negative $\tau$ (out of scope)  \\
Hezbollah action against Israel by Mar 20      & 2026-03-17  & 2026-03-02 & $-15.9$ & negative $\tau$ (out of scope)  \\
\midrule
2 additional markets                          & ---         & not recovered & --- & confidence below 0.7 threshold  \\
\bottomrule
\end{tabularx}
\end{table}

The dominant exclusion reason among recovered-$\Tevent$ markets is missing CLOB price coverage at the time of the evaluation (8 of 11 positive-$\tau$ markets). This reflects the Polymarket subgraph indexer's coverage policy: the indexer captures markets only after a transaction triggers indexing, and for low-volume markets the indexing latency can exceed the contract's full lifetime. Continuous CLOB price collection on all markets in the target categories from $\Topen$ onward is the natural remedy and is identified as the highest-priority infrastructure improvement in \Cref{sec:eval-refinements}.

\section{$\Tevent$ Recovery Verification Procedure}
\label{app:verification}

For each Tier 3 recovery used in the empirical evaluation we executed a two-stage verification. First, the LLM is required to cite at least three independent news sources for the recovered date. Second, the cited sources are cross-checked manually for date consistency and for the underlying event being plausibly the YES-resolving event of the contract. Disagreements among sources of more than 24 hours, or disagreements about whether the cited event qualifies under the contract's resolution criteria, downgrade the confidence to below the $0.7$ threshold and exclude the market from the empirical evaluation.

For the Iran-Apr30 market specifically, the recovered $\Tevent = $ April 3, 2026 was cross-verified across eight independent sources covering the F-15E special operations entry into Iran (Wikipedia, CBS News, Axios, TIME, Al Jazeera, NBC News, Washington Post, Reuters). The dates agreed within 12 hours across all sources; the qualifying-event test (does the F-15E operation constitute U.S.\ forces entering Iran?) was answered affirmatively by all sources that addressed the question explicitly. Confidence was set to $0.80$.

We note that this verification procedure does not establish that the recovered $\Tevent$ is the \emph{first} public observation of the event in the strict sense---earlier private knowledge, leaked but not yet widely reported, would not be detectable by this procedure. The verification establishes only that the recovered timestamp is the earliest publicly-reported date in the post-hoc news record. For the purposes of the deadline-ILS, this is the operationally relevant timestamp, since pre-event positioning is benchmarked against price movement before public knowledge becomes widely distributed.

\section*{Revision note (v2)}
\label{sec:revision-note}

This v2 revision updates the per-category hazard-rate estimates in \Cref{tab:hazard-fits} and Section~\ref{sec:eval-hazard} to reflect two findings from the platform data audit:

\textbf{(i) Esports taxonomy correction.} The v1 \texttt{military\_geopolitics} category inadvertently included $15{,}542$ esports markets (predominantly Counter-Strike match outcomes) due to keyword overlap on the word ``strike.'' These have been reclassified to a new \texttt{esports} category in the v1.1 taxonomy. Zero esports markets were in the Tier-3 hazard sample, so the reclassification has no effect on the hazard estimates themselves; the correction is reported for transparency in derived datasets.

\textbf{(ii) Tier-3 sample expansion.} The v1 hazard estimates were computed under a budget-capped Tier-3 sample of $n = 9$ markets for \texttt{military\_geopolitics}. The v2 estimates use the full Tier-3 population of $n = 18$ markets available at snapshot 2026-04-30. The point estimate moves from $\hat\lambda = 0.306$ (v1) to $\hat\lambda = 0.241$ (v2), a $21\%$ relative change driven entirely by the doubling of the sample. The v1 estimate lies inside the v2 95\% confidence interval $[0.143, 0.365]$; the two estimates are statistically consistent. The half-life moves from $2.3$ days (v1) to $2.9$ days (v2). KS adequacy is preserved ($p = 0.426$ in v2, $p = 0.609$ in v1; both exceed the $\alpha = 0.05$ threshold).

The substantive conclusions of v1 are unchanged. The fast-event-occurrence finding for military / geopolitical markets is confirmed by the larger sample; the bimodal-distribution finding for regulatory-decision markets is preserved (now $n = 16$, $p = 0.023$); the preliminary corporate-disclosure fit ($n = 5$) is reported as in v1. The single-case Iran-Apr30 $\ILS^{\text{dl}}$ computation ($+0.113$ vs proxy $-0.331$) is not affected by the hazard correction. The methodology refinements identified in v1 remain the work programme.

Three accompanying datasets are released openly with this revision: the $2{,}052$-market T\_news / T\_event recovery corpus underlying this paper's Tier 3 anchors (\textit{polymarket-tnews-tevent-recovery-v1}); the per-category hazard fits with full numerical detail (\textit{polymarket-hazard-rates-v1}); and the population-scale ILS for $4{,}801$ resolved markets (\textit{polymarket-ils-corpus-v1}). All three are available at \url{https://github.com/ForesightFlow/datasets}.

\bibliographystyle{plainnat}
\bibliography{refs}

\end{document}